\documentclass[conference]{IEEEtran}
\IEEEoverridecommandlockouts
\usepackage{cite}
\usepackage{amsmath,amssymb,amsfonts}
\usepackage{algorithmic}
\usepackage{graphicx}
\usepackage{textcomp}
\usepackage{xcolor}
\usepackage{numprint} 
\usepackage{caption}
\usepackage{multirow}
\usepackage{float}
\usepackage{tabularx}
\usepackage[table]{xcolor}
\usepackage{caption}
\usepackage[normalem]{ulem}

\renewcommand{\figurename}{Figure}
\usepackage[bookmarksnumbered, colorlinks, plainpages, linkcolor=black, citecolor=black, urlcolor=black]{hyperref}
\usepackage{tikz,xcolor,hyperref}
\definecolor{lime}{HTML}{A6CE39}
\DeclareRobustCommand{\orcidicon}{
	\begin{tikzpicture}
	\draw[lime, fill=lime] (0,0) 
	circle [radius=0.16] 
	node[white] {{\fontfamily{qag}\selectfont \tiny ID}};
	\draw[white, fill=white] (-0.0625,0.095) 
	circle [radius=0.007];
	\end{tikzpicture}
	\hspace{-2mm}
}
\foreach \x in {A, ..., Z}{\expandafter\xdef\csname orcid\x\endcsname{\noexpand\href{https://orcid.org/\csname orcidauthor\x\endcsname}
			{\noexpand\orcidicon}}
}

\usepackage[letterpaper, top=72pt, bottom=54pt, left=54pt, right=54pt]{geometry}
\def\BibTeX{{\rm B\kern-.05em{\sc i\kern-.025em b}\kern-.08em
    T\kern-.1667em\lower.7ex\hbox{E}\kern-.125emX}}
\begin{document}

\title{Machine Learning-Based Performance Evaluation of a Solar-Powered Hydrogen Fuel Cell Hybrid in a Radio-Controlled Electric Vehicle}

\newcommand{\orcidauthorA}{0009-0007-2690-4496}
\newcommand{\orcidauthorB}{0000-0002-9721-6291}
\newcommand{\orcidauthorC}{0000-0002-3250-0742}
\newcommand{\orcidauthorD}{0000-0002-5211-3598}

\author{
  Amirhesam Aghanouri~\orcidA{}, \emph{Member, IEEE},
  Mohamed Sabry~\orcidB{}, \emph{Member, IEEE},\\
  Joshua Cherian Varughese~\orcidC{}, \emph{Member, IEEE},
  Cristina Olaverri-Monreal~\orcidD{}, \emph{Senior Member, IEEE}%
  \thanks{Johannes Kepler University Linz, Austria; Department Intelligent Transport Systems.
  \texttt{\{amirhesam.aghanouri, mohamed.sabry, joshua.varughese, cristina.olaverri-monreal\}@jku.at}}
}

\maketitle
\begin{abstract}
This paper presents an experimental investigation and performance evaluation of a hybrid electric radio-controlled car powered by a Nickel-Metal Hydride battery combined with a renewable Proton Exchange Membrane Fuel Cell system. The study evaluates the performance of the system under various load-carrying scenarios and varying environmental conditions, simulating real-world operating conditions including throttle operation. In order to build a predictive model, gather operational insights, and detect anomalies, data-driven analyses using signal processing and modern machine learning techniques were employed. Specifically, machine learning techniques were used to distinguish throttle levels with high precision based on the operational data. Anomaly and change point detection methods enhanced voltage stability, resulting in fewer critical faults in the hybrid system compared to battery-only operation. Temporal Convolutional Networks were effectively employed to predict voltage behavior, demonstrating potential for use in planning the locations of fueling or charging stations. Moreover, integration with a solar-powered electrolyzer confirmed the system’s potential for off-grid, renewable hydrogen use. The results indicate that integrating a Proton Exchange Membrane Fuel Cell with Nickel-Metal Hydride batteries significantly improves electrical performance and reliability for small electric vehicles, and these findings can be a potential baseline for scaling up to larger vehicles.
\end{abstract}


\section{Introduction}
\label{sec:Introduction}
Battery-equipped vehicles can help reduce transport-related greenhouse gas emissions by relying partially or entirely on electric energy instead of fossil fuels \cite{khiari2022uncertainty, validi2021analysis}. In particular, the use of green hydrogen, produced via electrolysis using renewable energy sources such as solar power, offers a sustainable and zero-emission alternative for hydrogen-powered vehicles like Fuel Cell Electric Vehicles (FCEVs), where fuel cells generate electricity through an electrochemical process that emits only water vapor. This efficient, quiet, and low-emission technology is highly adaptable and can be integrated into vehicles of various scales, including Radio-Controlled (RC) cars. Using an RC car as an experimental platform allows for a small-scale rapid, safe, and cost-effective investigation of hydrogen fuel cell performance under diverse real-world conditions, without the logistical and regulatory complexities of full-scale vehicle testing.

We investigated in this paper the use of machine learning to model and enhance the operational efficiency of small-scale hydrogen fuel cell systems under real-world conditions. To this end, we developed and tested a hybrid electric RC vehicle platform, powered by a Nickel-Metal Hydride (NiMH) battery and a Proton Exchange Membrane Fuel Cell (PEMFC), under varying load and environmental scenarios. The experimental framework covers a wide range of idle and dynamic operating scenarios, including variable throttle levels, loads, surface conditions, and ambient temperatures, both indoors and outdoors. Hydrogen for the fuel cell is produced via grid-powered and solar-powered Proton Exchange Membrane (PEM) electrolysis, demonstrating both practical and renewable refueling options.

High-resolution data from the RC test platform is collected using a portable sensor suite, processed through signal filtering, and analyzed via supervised learning for throttle classification, anomaly and change point detection, and deep learning (Temporal Convolutional Network (TCN)) for predictive modeling.

The main contributions of this work are:

\begin{itemize}
    \item Development of a portable, scenario-adaptive sensor suite for data acquisition.
    \item Application of advanced data-driven techniques such as signal processing, supervised classification, anomaly detection, and deep learning for operational monitoring and predictive modeling of small-scale hydrogen fuel cell systems.
\end{itemize}

This paper is structured as follows: Section \ref{sec:Related literature} shows an overview of related research on hydrogen fuel cells in small-scale vehicles and advanced analytical methods.
Section \ref{sec:Methodology} presents the experimental setup, data collection, and the analysis approach. The experimental results are presented in Section \ref{sec:Results}. Finally, conclusions and future work are presented in Section \ref{sec:Conclusion and Future work}.
 
\section{Related literature}
\label{sec:Related literature}
Many studies have investigated the electrical and kinematics of RC cars, however, the number of works that specifically test RC cars powered by fuel cells remains limited. In \cite{mohamad2007overview}, the authors perform one of the first studies that gives an overview of an RC car retrofitted with a fuel cell, describing its design and how it powers the motor and electronics. The aim is to show fuel cells as a cleaner option for small vehicles. However, it lacks detailed data and detailed performance analysis, serving mainly as a conceptual demonstration. Another study introduced the generation of hydrogen from waste aluminum to power a fuel cell in an RC car, demonstrating a practical way to recycle aluminum for clean energy. The authors develop a mathematical model to predict hydrogen production from aluminum-water reactions and tested this model. The work focuses mainly on the chemical aspects of hydrogen production and provides only limited insights into the detailed operational behavior of the RC car \cite{saluena2021waste}.

A further study presented the design and testing of an electric RC vehicle powered exclusively by a proton exchange membrane fuel cell stack, using a metal hydride tank for hydrogen storage and electrolytic capacitors to supply peak currents instead of conventional batteries \cite{beneito2007electric}. Their prototype demonstrated the practical feasibility of replacing batteries in small electric vehicles with a cleaner hydrogen-based system. However, this approach showed clear limitations, including insufficient hydrogen flow due to inadequate thermal management, and the absence of automated control and monitoring, which constrain its practicality and scalability. Another work on hydrogen fuel cell trucks has shown the feasibility of using PEMFC, metal hydride storage, and super capacitor buffers, with a small-scale model and wireless data logging \cite{misiopecki2011investigation}. However the experimental validation was not provided. Thermal management lacked active control for stable operation since it depended on ambient heat and exhaust. Additionally, no dynamic experiments or statistical trend analysis were conducted, and the data acquisition system only logged data without real-time processing or predictive modeling.

Although previous studies have demonstrated the feasibility of using fuel cells in RC vehicles, few have integrated comprehensive sensor systems, data acquisition across diverse scenarios, or applied machine learning for comprehensive performance analysis. Therefore, this study addresses these gaps by developing and testing a 1/10-scale RC car platform powered by a hybrid system combining a NiMH battery with a PEMFC module. The platform provides active sensing, real-time monitoring, and renewable hydrogen generation, addressing critical challenges in thermal management, system control, and predictive modeling identified in prior research.

\section{Methodology}
\label{sec:Methodology}
This section presents the methodology which was implemented to process and analyze experimental data collected from an RC platform under diverse conditions.

\subsection{Hydrogen generation}
\begin{figure}[ht]
    \centering
    \includegraphics[width=0.43\textwidth]{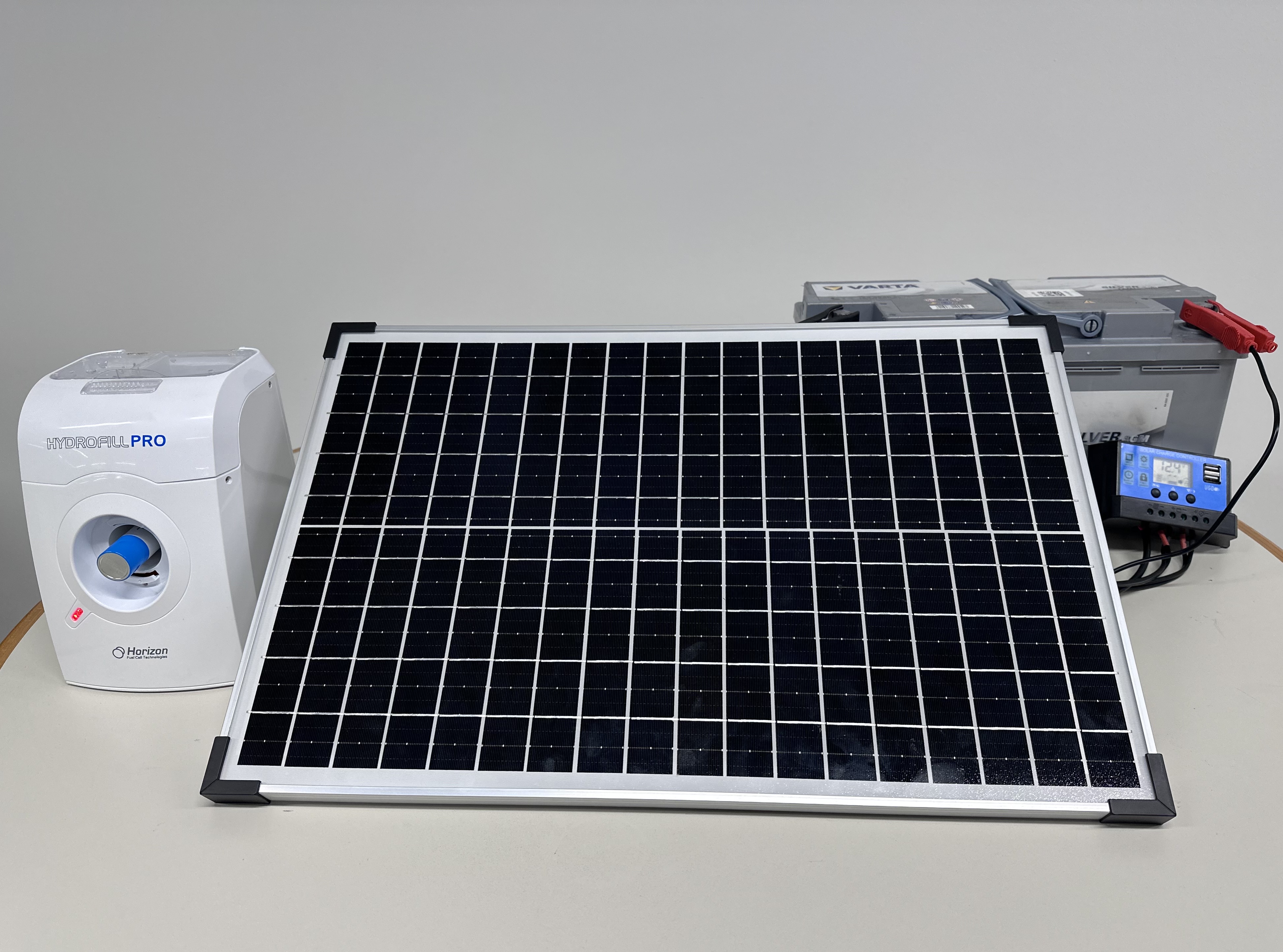}
    \caption{Experimental setup for renewable hydrogen generation, including a hydrogen generator, a 30 $W$ monocrystalline silicon solar panel, a charge controller, and lead-acid batteries. This configuration enables safe and repeatable off-grid hydrogen production by using clean solar energy to electrolyze deionized water and refill hydrogen cartridges sustainably and efficiently.}
    \label{fig:solar}
\end{figure}

Hydrogen generation for the cartridges, as shown in \autoref{fig:solar}, was carried out using two complementary approaches to ensure safe, repeatable, and sustainable operation. The first involved generating hydrogen through PEM electrolysis of deionized water, operating with a direct current input of 10–19 $V$ and up to 23 $W$ of power. Up to 3 $L/h$ of hydrogen was produced while approximately 20 $ml/h$ of water was consumed. Each cartridge was typically refilled in approximately 4 to 5 hours at 25 °C using electricity. The second approach involved powering the hydrogen generator with a 30 $W$ monocrystalline silicon solar panel (Vmp = 18 $V$, Imp = 1.67 $A$, 530 mm × 350 mm × 17 mm), to demonstrate a practical, renewable, and fully off-grid hydrogen supply. Using solar energy extended the refill time to about 5–7 hours, depending on weather conditions.

\subsection{Experimental setup}
The tests were carried out at multiple throttle levels in both idle and dynamic scenarios, covering indoor and outdoor environments with varying temperatures, surface types, and different loading and towing configurations. All scenarios were tested in two configurations: the first used a single Ni-MH battery pack (7.2 $V$ nominal, 4000 $mAh$), while the second combined the same battery with a 30 $W$ PEMFC module. NiMH battery tests were performed at 25\%, 50\%, 75\%, and 100\% throttle in idle conditions for 1 minute each indoors without load. Dynamic tests included driving inside the lab for 10 minutes without load, driving with 1 $kg$ load for 10 minutes, towing 3 $kg$ for 5 minutes indoors, and outdoor asphalt driving for 5 minutes without load under varying temperatures. The same scenarios were repeated with the PEMFC fuel cell added to supplement the battery, providing more consistent power.

\begin{figure}[ht]
    \centering
    \includegraphics[width=0.44\textwidth]{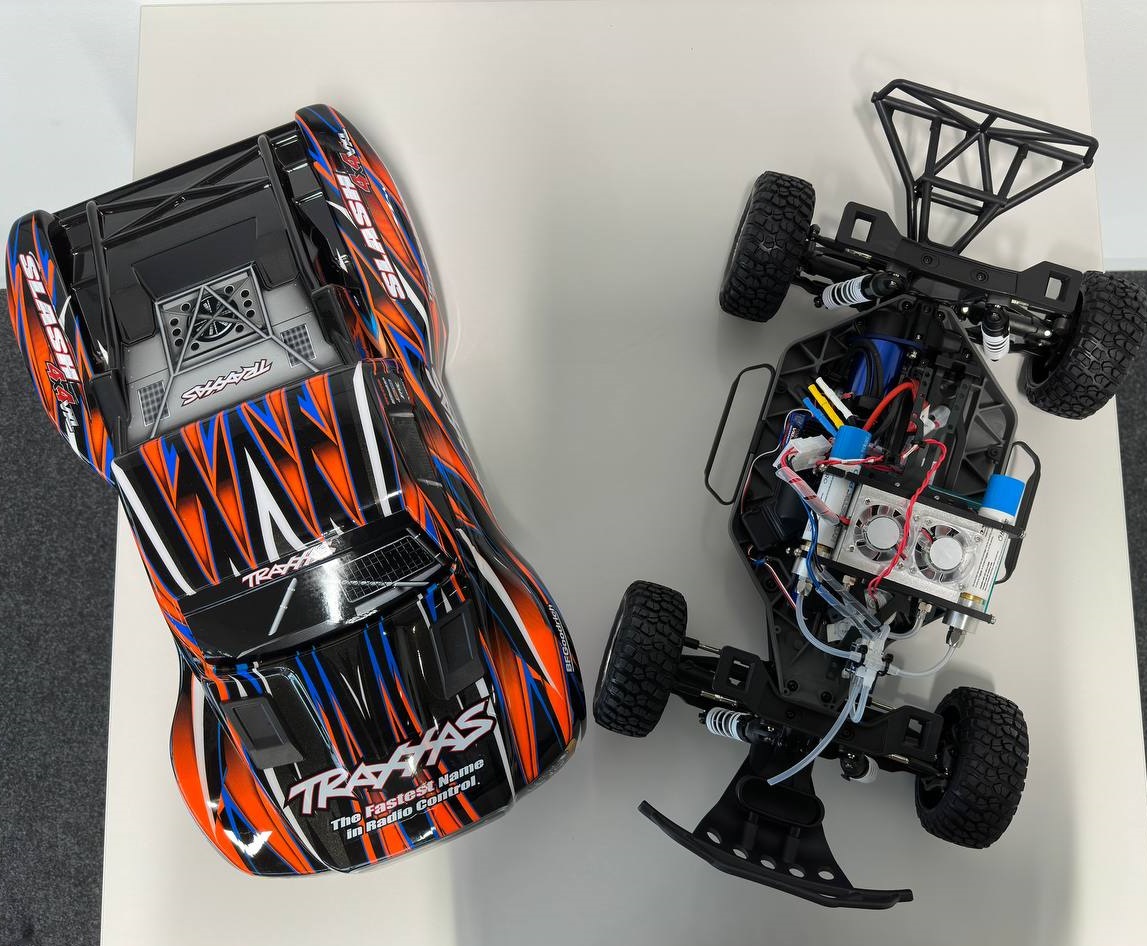}
    \caption{The Traxxas Slash 4×4 VXL 1/10-scale RC test platform used for hybrid powertrain experiments with a NiMH battery and a 30 $W$ PEM hydrogen fuel cell module.}
    \label{fig:Hybrid_car}
\end{figure}

A Traxxas Slash 4×4 VXL, a 1/10-scale 4WD RC car, as shown in \autoref{fig:Hybrid_car}, with a brushless power system, was used for the experiments. Based on the configurations, the RC car reaches a maximum speed of approximately 35 to 40 mph. In the case of using the PEMFC module, the hydrogen for the fuel cell was stored in two metal hydride cartridges containing an $AB_2$ alloy within an aluminum casing to enhance heat transfer. Each cartridge, equipped with integrated valves, regulators, and electronic controls, operates at approximately 30 bar and stores up to 10 $L$ of high-purity hydrogen, ensuring safe discharge while preventing leakage or backflow.

\subsection{Data acquisition and sensor integration}
To facilitate comprehensive data acquisition and environmental monitoring, an Arduino Mega 2560 microcontroller was interfaced with a suite of external sensors. All signals were acquired at a sampling frequency of about 8.25 $Hz$. A brief description of the setup is listed below.
\begin{itemize}
\item Voltage sensor: Measures voltage fluctuations supplied to the electrical motor of the RC car to monitor system voltage.

\item Current sensor: Measures the current flowing to the motor by sensing the current drawn from the battery/hybrid unit.

\item BMP180 sensor: Acquires environmental data, including altitude, atmospheric pressure, and temperature.

\item SD card: Enables autonomous data logging on the platform and manages the recording and saving of data from multiple scenarios for efficient retrieval.
\end{itemize}

\subsection{Data processing and signal filtering}
As voltage, current, and power measurements are affected by random fluctuations due to sensor limitations and environmental factors, a Simple Moving Average (SMA) filter was applied to each signal to suppress noise while preserving meaningful variations. Smoothing provides the ability to reduce high-frequency noise and reveal the true system behavior, helping to assess the impact of using a PEMFC on the car’s performance. The SMA window size was adjusted based on the test duration to balance noise reduction with temporal resolution. To better illustrate signal trends, the raw signals were visualized alongside their SMA to highlight underlying patterns, as described in the next section. Moreover, the correlations between voltage, current, power, and temperature were examined to understand how the signals relate to each other and how these relations change under different operating scenarios.

\subsection{Machine learning analysis}
\subsubsection{Supervised classification of throttle percentages}
To classify throttle percentages under stationary test conditions, a supervised machine learning framework was developed. Input features included voltage, current, temperature, and derived power metrics, along with a system configuration indicator. The target variable, throttle percentage, was encoded as a multi-class label. Two classifiers, including Random Forest (RF) \cite{breiman2001random} and Gradient Boosting (GB) \cite{friedman2001greedy}, were chosen for their robustness to noise, ability to capture nonlinear feature interactions, and interpretability via feature importance. Both classifiers were trained and tuned via grid search to optimize hyperparameters. Models were evaluated using accuracy scores and confusion matrices to compare predicted and true throttle classes. This analysis was conducted to determine whether electrical and thermal measurements alone could be used to distinguish throttle levels with high confidence, which is critical for condition monitoring and control applications.

\subsubsection{Anomaly and change point detection}
To objectively identify outliers and shifts in the electrical signals, automated anomaly and change point detection method were applied. The modified Z-score method \cite{iglewicz1993volume} was used to detect individual outliers in voltage and current. Unlike the traditional Z-score, the modified Z-score relies on the median and Median Absolute Deviation (MAD) instead of the mean and standard deviation, making it more robust and less sensitive to extreme values. A data point is flagged as an anomaly if its modified Z-score exceeds a threshold, which was set to 3 in this study, ensuring that outliers are accurately detected even in the presence of skewed or heavy-tailed distributions. Furthermore, the Pruned Exact Linear Time (PELT) algorithm \cite{killick2012optimal} was utilized to locate structural change points in the signals. These combined methods ensured consistent, quantitative detection of anomalies and operational transitions, providing a more reliable basis for interpreting system behavior under various driving conditions.

\subsection{Temporal convolutional networks}
To model and predict the system’s time-dependent behavior under various operating scenarios, a temporal convolutional network \cite{bai2018empirical} was implemented. The TCN architecture uses stacked one-dimensional convolutional layers with causal and dilated convolutions to capture both short and long-range temporal dependencies. Voltage signals were first normalized using a min-max scaling approach, then structured into overlapping input sequences. A grid search was performed to fine-tune key hyperparameters, including sequence length, number of filters, kernel size, number of layers, dropout rate, and batch size, ensuring optimal model performance. The trained TCN predicts future values of the electrical signals, enabling assessment of the car’s dynamic response and powertrain behavior with and without the PEMFC. The model was evaluated under four representative dynamic indoor driving scenarios: driving the RC car in battery-only mode and hybrid mode, both with and without an additional 1 $kg$ load. Predicted outputs were compared with true measurements to evaluate the model’s accuracy and to demonstrate the potential of TCNs for robust, data-driven time-series prediction in small-scale hybrid electric vehicles.

The proposed voltage prediction model was evaluated under four representative dynamic indoor driving scenarios where two of them are for driving without and with 1 $kg$ load powering the RC Car bz battery only and same secarios with hybrid where PEMFC module was cosidered, comparing a battery-only configuration with a hybrid setup, both with and without an added load.

\section{Results}
\label{sec:Results}
This section presents the main findings from the experiments, highlighting how the hybrid fuel cell system affected the RC car’s performance under different scenarios.

\subsection{Results from the comparison of battery-only vs hybrid configuration}
\label{sec:results_SMA}
\begin{figure}[htb]
    \centering
    \includegraphics[width=0.48\textwidth]{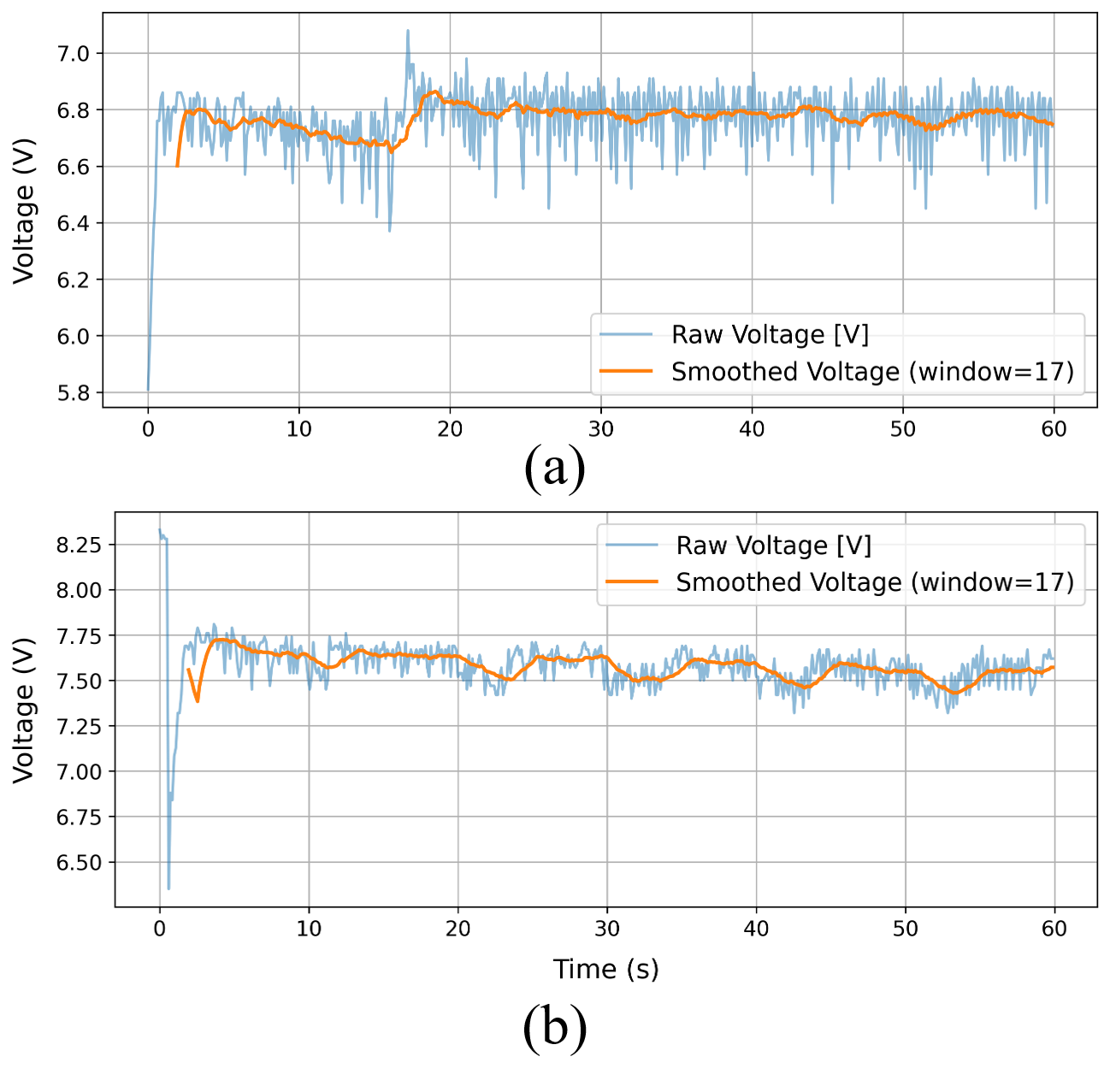}
    \caption{Voltage response at 100\% throttle under static conditions: Figure (a) battery-only shows greater voltage drop and fluctuations; (b) The hybrid configuration consistently maintained a more stable voltage profile, as demonstrated by data smoothed using the moving average filter with the window size of 17.}
    \label{fig:V_static_100_throttle_a_bat_b_hyb}
\end{figure}

The results shown in \autoref{fig:V_static_100_throttle_a_bat_b_hyb} compare the static 100\% throttle performance for one minute of the battery-only and hybrid configurations without mechanical load. A moving average filter with a window size of 17 samples (2 seconds) was applied to the voltage data. As illustrated, the hybrid configuration maintained a more stable voltage, whereas the battery-only setup experienced a larger initial voltage drop and greater fluctuations throughout the test, with the hybrid system demonstrating improved voltage regulation and reduced transient noise under sustained high-current demand.

\begin{figure}[htb]
    \centering
    \includegraphics[width=0.47\textwidth]{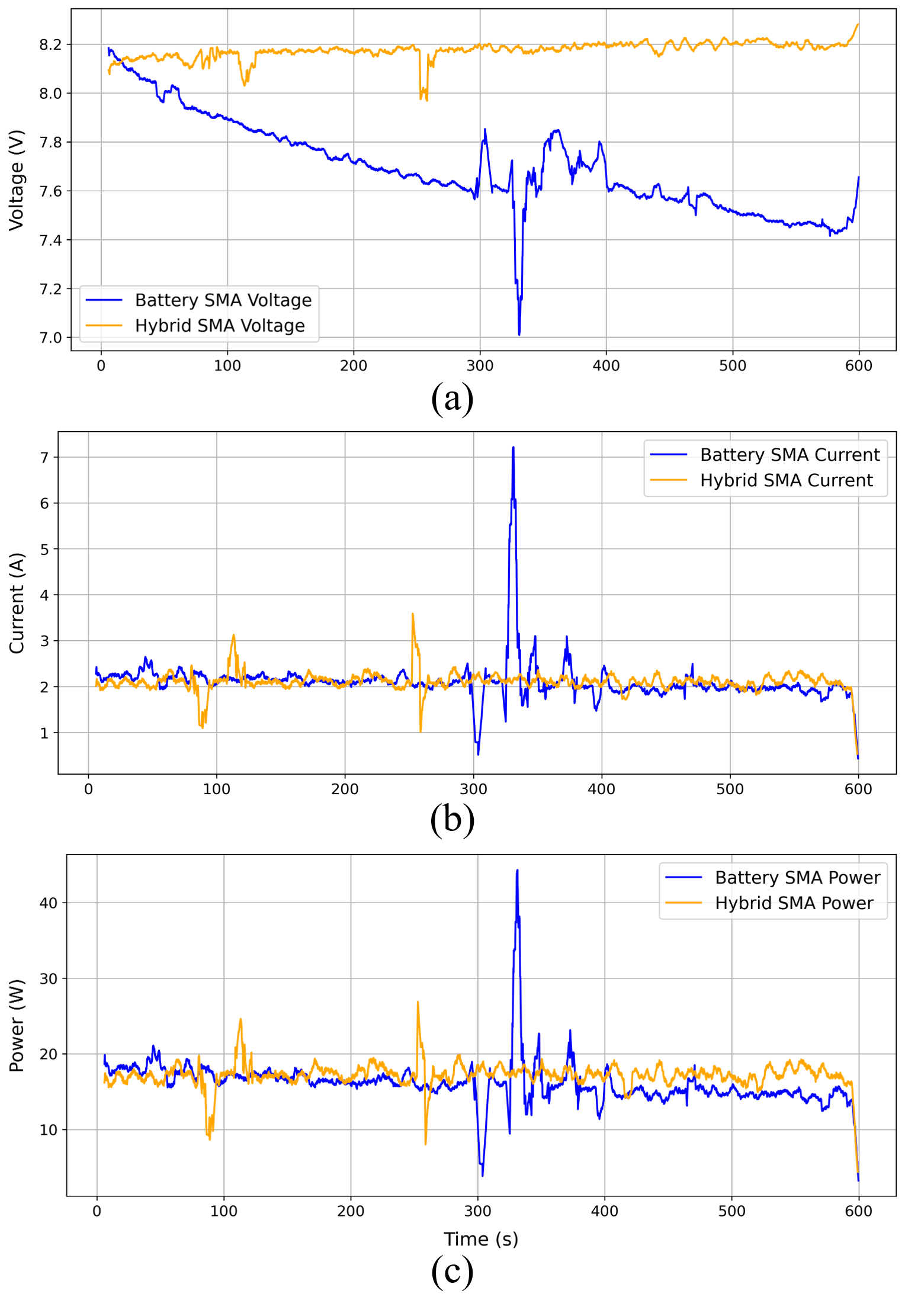}
    \caption{(a) Voltage, (b) current, and (c) power signals smoothed using SMA (window size 50) for battery-only and hybrid configurations during the 10-minute dynamic lab driving test without load. The hybrid system maintains higher voltage and smoother current and power profiles compared to the battery-only mode.}
    \label{fig:10_minutes_drive_no_load_signals_compared}
\end{figure} 

\autoref{fig:10_minutes_drive_no_load_signals_compared} shows the SMA with a window size of 50 for voltage, current, and power for battery-only and hybrid configurations during the 10-minute dynamic driving test without load. The voltage profiles indicated that the hybrid system consistently maintained a higher terminal voltage than the battery-only mode. While the battery-only voltage gradually declined due to continuous discharge, the hybrid configuration kept the voltage more stable, suggesting that the fuel cell supplemented the load and partially recharged the battery during operation. The current and power plots also show that the hybrid mode smoothed out fluctuations and peak demands compared to the battery alone, which experienced higher variability, resulting in a comparatively steadier performance. 

\autoref{tab:correlation_matrices} presents the correlation matrices for the towing and 1 $kg$ load scenarios under battery-only and hybrid configurations. The hybrid system exhibited weaker negative correlations between V–I and V-P compared to the battery-only case, indicating reduced voltage fluctuations under higher load conditions. The consistently strong I–P correlations across all scenarios confirmed the expected dominant influence of current on power output. Moreover, the V-T correlation shifted from strongly positive values in the battery-only setup to higher or even negative values in the hybrid system, suggesting that the integration of the fuel cell helped in buffering thermal effects during prolonged operation. Furthermore, in no-load driving scenarios lasting 5 minutes inside and outside the lab, with surface changes and a temperature rise of about 10\textit{°C}, the fuel cell notably reduced voltage sensitivity to temperature indoors, stabilizing power despite the changes. Outdoors, where conditions naturally support battery stability, the fuel cell showed only a minor increase in temperature-related voltage variation.

\begin{table}[htbp]
\centering
\caption{Correlation matrices for towing and 1\,$kg$ load scenarios (battery-only vs hybrid).}
\label{tab:correlation_matrices}
\small
\begin{tabular}{lcccc}
\hline
\textbf{Scenario} & \textbf{V--I} & \textbf{V--P} & \textbf{I--P} & \textbf{V--T} \\
\hline
Towing (Battery) & -0.381 & -0.306 & 0.993 & 0.450 \\
Towing (Hybrid) & -0.175 & -0.100 & 0.996 & 0.544 \\
1\,$kg$ (Battery) & -0.188 & -0.122 & 0.997 & 0.737 \\
1\,$kg$ (Hybrid) & -0.143 & -0.111 & 0.999 & -0.109 \\
No load indoor (Battery) & 0.055 & 0.147 & 0.991 & 0.694 \\
No load indoor (Hybrid) & -0.313 & -0.246 & 0.995 & -0.090 \\
No load outdoor (Battery) & -0.267 & -0.133 & 0.990 & 0.064 \\
No load outdoor (Hybrid) & -0.189 & -0.119 & 0.997 & 0.245 \\
\hline
\end{tabular}

\vspace{4pt}
\footnotesize
\textbf{Abbreviations:} V = Voltage, I = Current, P = Power, T = Temperature.
\renewcommand{\arraystretch}{1} 
\end{table}

\subsection{Throttle classification results}
\label{sec:throttle_results}
With the best hyperparameters, the RF classifier achieved its best performance with 50 estimators, a maximum tree depth of 10, and a minimum samples split of 5. The GB classifier performed optimally with 100 estimators, a learning rate of 0.1, and a maximum depth of 3. \autoref{tab:classification_metrics} shows both classifiers achieved the highest precision and recall for the 25\% throttle level, near 0.98, while the 100\% throttle level was the most challenging, with precision and recall dropping to around 0.70–0.74. Intermediate throttle levels showed moderate classification performance, reflecting some confusion between neighboring classes. This result showed that understanding model reliability across throttle settings can help optimize control strategies in practical applications.

\begin{table}[htbp]
\centering
\caption{Evaluation metrics for classifier-based throttle percentage prediction.}
\label{tab:classification_metrics}
\small
\begin{tabular}{clcccc}
\hline
\textbf{Classifier} & \textbf{Metric} & \textbf{25\%} & \textbf{50\%} & \textbf{75\%} & \textbf{100\%} \\
\hline
\multirow{3}{*}{RF} 
 & Precision & 0.98 & 0.85 & 0.76 & 0.73 \\
 & Recall    & 0.97 & 0.84 & 0.81 & 0.70 \\
 & F1-score  & 0.98 & 0.85 & 0.78 & 0.72 \\
\hline
\multirow{3}{*}{GB} 
 & Precision & 0.98 & 0.89 & 0.75 & 0.74 \\
 & Recall    & 0.98 & 0.85 & 0.81 & 0.69 \\
 & F1-score  & 0.98 & 0.87 & 0.78 & 0.71 \\
\hline
\end{tabular}
\end{table}

\subsection{Analysis outcomes for anomaly and change point detection}
\begin{figure}[htb]
    \centering
    \includegraphics[width=0.48\textwidth]{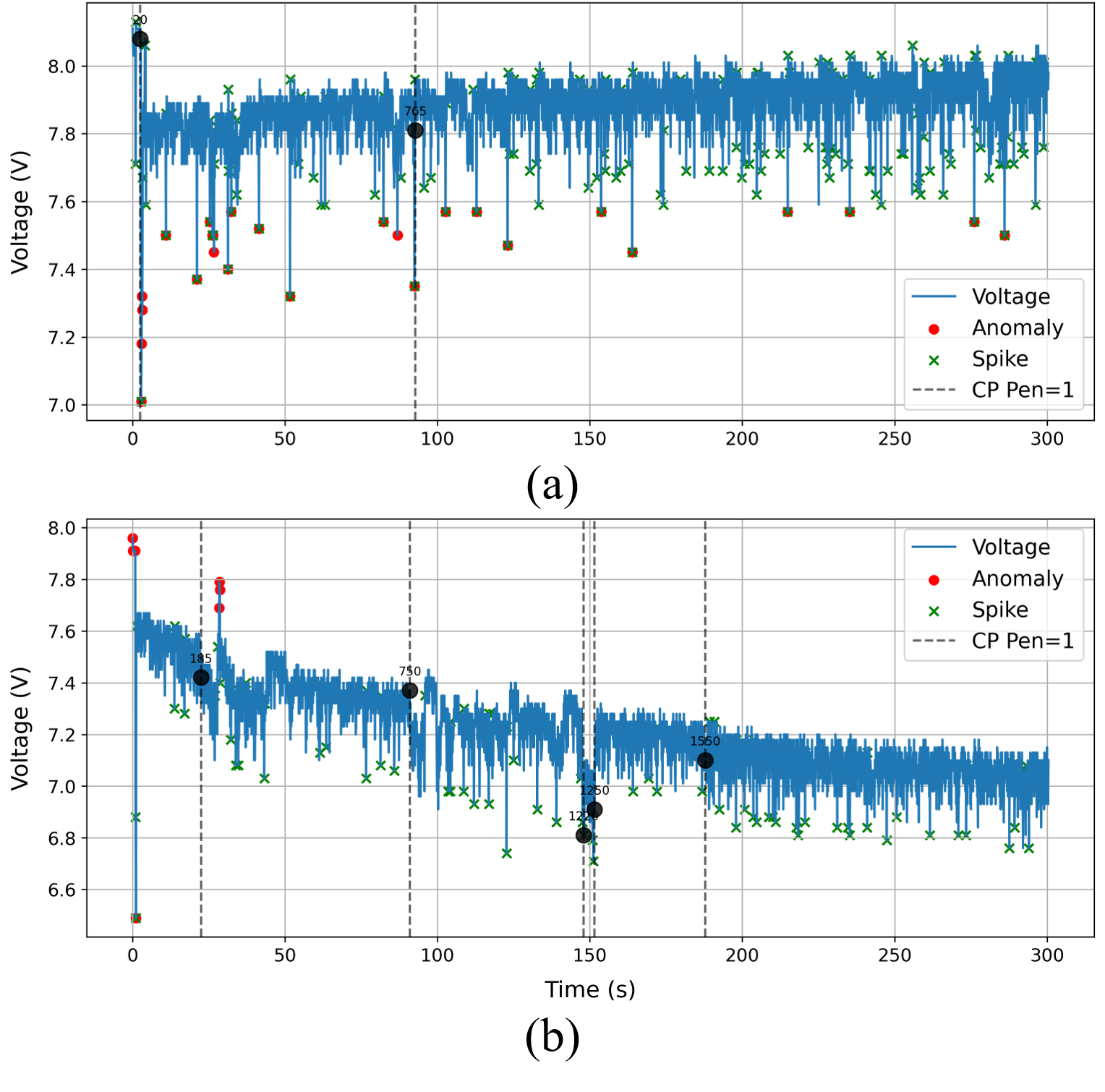}
    \caption{Voltage anomaly and change point detection results: (a) hybrid system and (b) battery-only under a load. The hybrid setup shows enhanced voltage stability with fewer structural changes and more frequent but minor anomalies, while the battery-only configuration exhibits fewer anomalies but suffers from a steady voltage decline, indicating progressive battery depletion.}
    \label{fig:anomaly_results}
\end{figure}

The results in \autoref{fig:anomaly_results} clearly demonstrate that integrating a PEMFC with the battery significantly enhanced voltage stability under load. While the hybrid system exhibited minor voltage anomalies (26 anomalies, spike threshold 0.2067 $V$) due to dynamic load-sharing between the fuel cell and battery, it maintained a higher and steadier average voltage, with only two structural change points detected by the PELT algorithm. In contrast, the battery-only configuration showed fewer isolated anomalies (12 anomalies,higher spike threshold 0.2406 $V$) but suffered from a clear voltage decline, confirmed by five structural change points indicating progressive depletion. Therefore, for sustained operation under load, the hybrid configuration offered superior electrical stability and reliability compared to the battery-only system.

\subsection{TCN prediction performance}
\begin{figure}[htb]
    \centering
    \includegraphics[width=0.49\textwidth]{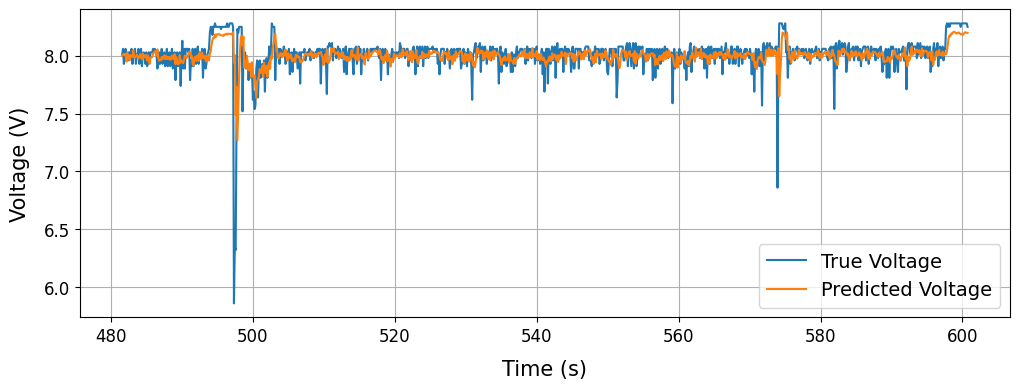}
    \caption{Comparison of true and predicted voltage for the hybrid configuration under a 1 $kg$ load during dynamic indoor driving for 10 minutes. The plot illustrates the model's ability to track voltage variations despite the added load and increased dynamic power demands, demonstrating that prediction errors remain within acceptable limits for real-time energy management.}
    \label{fig:voltage_prediction_1kg_hybrid}
\end{figure}

Voltage prediction model results under indoor driving scenarios (as described in \autoref{sec:Methodology}) demonstrated consistently low prediction errors across all conditions (see \autoref{tab:voltage_prediction}). Adding the fuel cell generally improves prediction accuracy in the no-load scenario, indicating more stable voltage behavior when the battery is supplemented by the PEMFC module. However, under loaded conditions, the increased dynamic power demands introduced greater fluctuations, slightly raising the prediction error for the hybrid configuration. The best hyperparameter configuration found through grid search was as follows: sequence length of 40 time steps, 64 convolutional filters, a kernel size of 2, 2 convolutional layers, and a batch size of 16. This configuration was used for the final prediction results. \autoref{fig:voltage_prediction_1kg_hybrid} shows the voltage prediction for the hybrid scenario with a 1 $kg$ load.

\begin{table}[htbp]
\centering
\caption{Voltage prediction performance for dynamic indoor driving scenarios.}
\label{tab:voltage_prediction}
\small
\begin{tabular}{ccc}
\hline
\textbf{Scenario} & \textbf{MAE} & \textbf{RMSE} \\
\hline
No load indoor (Battery) & 0.0641 & 0.0892 \\
1~kg load indoor (Battery) & 0.0548 & 0.0867 \\
No load indoor (Hybrid) & 0.0404 & 0.0623 \\
1~kg load indoor (Hybrid) & 0.0712 & 0.1384 \\
\hline
\end{tabular}
\end{table}

\section{Conclusion and future work}
\label{sec:Conclusion and Future work}
This study demonstrates that integrating a PEMFC with a NiMH battery significantly enhances the electrical performance, reliability, and runtime of a hybrid RC vehicle under dynamic and variable load conditions. The hybrid configuration provided key advantages for applications requiring consistent, high-quality power output, such as improved voltage stability, reduced battery stress, smoother power delivery, and enhanced thermal management compared to battery-only operation. Reduced battery stress also lowered the risk of failure, increasing the viability of such systems for scale-up to full-sized electric vehicles.

A portable sensor suite was used to collect high-resolution electrical and thermal data during testing across multiple driving scenarios. Signal processing and machine learning techniques (including supervised classification and anomaly detection) enabled real-time monitoring and precise inference of throttle levels without direct input. This approach supported robust diagnostics, control, and predictive maintenance in small hybrid electric platforms. TCNs reliably predicted voltage trends across varying loads and throttle inputs, showcasing their potential for intelligent energy management.

Anomaly detection methods confirmed the hybrid system's superior fault tolerance and power consistency, making it suitable for use in autonomous robots, off-grid systems, and other mission-critical applications. Furthermore, the use of a solar-powered electrolyzer for hydrogen generation validated the feasibility of portable, emission-free refueling without reliance on compressed gas or grid electricity, reinforcing the sustainability of the system.

The proposed hybrid platform and its integrated monitoring framework offer a foundation for developing smart control strategies to optimize energy use and extend system life. Future extensions may include the integration of perception sensors, enabling autonomous functions such as obstacle avoidance and intelligent navigation, and further demonstrating the potential of hydrogen-powered systems in self-driving and renewable mobility applications.

\section{Acknowledgement}
\label{sec:Acknowledgement}
This work was supported through the GH2MOB, Applicability And Limitations Of Green Hydrogen For Smart Local Mobility, funded project by Lift-C with the support of Upper Austria. Project number: LF0112451001

\bibliographystyle{IEEEtran}
\bibliography{bliblio}

\begin{thebibliography}{10}
\providecommand{\url}[1]{#1}
\csname url@samestyle\endcsname
\providecommand{\newblock}{\relax}
\providecommand{\bibinfo}[2]{#2}
\providecommand{\BIBentrySTDinterwordspacing}{\spaceskip=0pt\relax}
\providecommand{\BIBentryALTinterwordstretchfactor}{4}
\providecommand{\BIBentryALTinterwordspacing}{\spaceskip=\fontdimen2\font plus
\BIBentryALTinterwordstretchfactor\fontdimen3\font minus \fontdimen4\font\relax}
\providecommand{\BIBforeignlanguage}[2]{{%
\expandafter\ifx\csname l@#1\endcsname\relax
\typeout{** WARNING: IEEEtran.bst: No hyphenation pattern has been}%
\typeout{** loaded for the language `#1'. Using the pattern for}%
\typeout{** the default language instead.}%
\else
\language=\csname l@#1\endcsname
\fi
#2}}
\providecommand{\BIBdecl}{\relax}
\BIBdecl

\bibitem{khiari2022uncertainty}
J.~Khiari and C.~Olaverri-Monreal, ``Uncertainty-aware prediction of battery energy consumption for hybrid electric vehicles,'' in \emph{2022 IEEE Intelligent Vehicles Symposium (IV)}.\hskip 1em plus 0.5em minus 0.4em\relax IEEE, 2022, pp. 1005--1010.

\bibitem{validi2021analysis}
A.~Validi, W.~Morales-Alvarez, and C.~Olaverri-Monreal, ``Analysis of the battery energy estimation model in sumo compared with actual analysis of battery energy consumption,'' in \emph{2021 16th Iberian Conference on Information Systems and Technologies (CISTI)}.\hskip 1em plus 0.5em minus 0.4em\relax IEEE, 2021, pp. 1--6.

\bibitem{mohamad2007overview}
A.~Mohamad and N.~A. Rahim, ``Overview of fuel cell powered radio controlled car,'' in \emph{Engineering conference on energy \& environment (EnCon), Sarawak, Malaysia}, 2007, pp. 285--8.

\bibitem{saluena2021waste}
X.~Salue{\~n}a-Berna, M.~Mar{\'\i}n-Genesc{\`a}, L.~Massagu{\'e}s~Vidal, and J.~M. Dag{\`a}-Monmany, ``Waste aluminum application as energy valorization for hydrogen fuel cells for mobile low power machines applications,'' \emph{Materials}, vol.~14, no.~23, p. 7323, 2021.

\bibitem{beneito2007electric}
R.~Beneito, J.~Vilaplana, and S.~Gisbert, ``Electric toy vehicle powered by a pemfc stack,'' \emph{International journal of hydrogen energy}, vol.~32, no. 10-11, pp. 1554--1558, 2007.

\bibitem{misiopecki2011investigation}
C.~Misiopecki, ``Investigation of fuel cell technology for long-haul trucks,'' Ph.D. dissertation, 2011.

\bibitem{breiman2001random}
L.~Breiman, ``Random forests,'' \emph{Machine learning}, vol.~45, pp. 5--32, 2001.

\bibitem{friedman2001greedy}
J.~H. Friedman, ``Greedy function approximation: a gradient boosting machine,'' \emph{Annals of statistics}, pp. 1189--1232, 2001.

\bibitem{iglewicz1993volume}
B.~Iglewicz and D.~C. Hoaglin, \emph{Volume 16: how to detect and handle outliers}.\hskip 1em plus 0.5em minus 0.4em\relax Quality Press, 1993.

\bibitem{killick2012optimal}
R.~Killick, P.~Fearnhead, and I.~A. Eckley, ``Optimal detection of changepoints with a linear computational cost,'' \emph{Journal of the American Statistical Association}, vol. 107, no. 500, pp. 1590--1598, 2012.

\bibitem{bai2018empirical}
S.~Bai, J.~Z. Kolter, and V.~Koltun, ``An empirical evaluation of generic convolutional and recurrent networks for sequence modeling,'' \emph{arXiv preprint arXiv:1803.01271}, 2018.

\end{thebibliography}

\end{document}